\documentclass[10pt,conference,letterpaper]{IEEEtran}

\usepackage{times,amsmath}
\usepackage{amsfonts}
\usepackage{amssymb,bm}
\usepackage{amsthm}
\usepackage{algorithm}
\usepackage{algpseudocode}
\usepackage{cite}
\usepackage{graphicx}
\usepackage{epstopdf}
\usepackage[caption=false,font=footnotesize]{subfig}
\usepackage{fixltx2e}

\begin{document}
%
\title{Achieving Max-Min Fairness for MU-MISO with Partial CSIT: A Multicast Assisted Transmission}

\author{
\IEEEauthorblockN{Hamdi Joudeh\IEEEauthorrefmark{1} and Bruno Clerckx\IEEEauthorrefmark{1}\IEEEauthorrefmark{2}}
\fontsize{9}{9}\upshape
\IEEEauthorrefmark{1} Department of Electrical and Electronic Engineering, Imperial College London, United Kingdom \\
\IEEEauthorrefmark{2} School of Electrical Engineering, Korea University, Seoul, Korea \\
\fontsize{9}{9}\selectfont\ttfamily\upshape
Email: \{hamdi.joudeh10, b.clerckx\}@imperial.ac.uk
}

\maketitle

\begin{abstract}
We address the max-min fairness design problem for a MU-MISO system with partial Channel State Information (CSI) at the Base Station (BS), consisting of an imperfect channel estimate and statistical knowledge of the estimation error, and perfect CSI at the receivers.
The objective is to maximize the minimum Average Rate (AR) among users subject to a transmit power constraint.
An unconventional transmission scheme is adopted where the Base Station (BS) transmits a common message in addition to the conventional private messages.
In situations where the CSIT is not accurate enough to perform interference nulling, individual rates are assisted by allocating parts of the common message to different users according to their needs.
The AR  problem is transformed into an augmented Average Weighted Mean Square Error (AWMSE) problem, solved using Alternating Optimization (AO).
The benefits of incorporating the common message are demonstrated through simulations.
\end{abstract}


\IEEEpeerreviewmaketitle

\section{Introduction}
\newcounter{Proposition_Counter} 
\newcounter{Lemma_Counter} 
\newcounter{Remark_Counter} 
\newcounter{Assumption_Counter}
We consider a Multi-User (MU) Multiple Input Single Output (MISO) system where a Base Station (BS) equipped with $N_{t}$ antennas serves $K \leq N_{t}$ single-antenna receivers.
For a given transmit power $P_{t}$, the objective is to design precoders (or beamformers) that maximize the minimum rate among users;
a criteria known in literature as max-min fairness \cite{Wiesel2006,Karipidis2008,Razaviyayn2013b}.
In practical scenarios, perfect Channel State Information at the Transmitter (CSIT) is unguaranteed due to the varying nature of wireless channels and the limited resources available for CSIT acquisition.
This prompted a number of robust fairness-based designs \cite{Vucic2009a,Bogale2011,Joudeh2014}.
In this paper, we consider the case where the BS obtains an imperfect estimate of the channel (through feedback or training) and statistical properties of the estimation error, while users estimate and track their channels accurately.
The problem is formulated in terms of the Average Rate (AR) which captures the overall performance w.r.t channel estimation errors.
Contrary to our previous work in \cite{Joudeh2014} which considers a similar scenario, a rather unorthodox transmission scheme is considered in this paper.
In particular, the BS transmits a common message in addition to the conventionally transmitted private messages.
This scheme is termed Joint Multicasting and Broadcasting (JMB).

JMB was employed in \cite{Hao2013} to boost the achievable Degrees of Freedom (DoF) under estimation errors that decay with increased Signal to Noise Ratio (SNR).
Assuming fixed noise variances at the receivers, the decaying power of the estimation error scales as $O\big(P_{t}^{-\alpha}\big)$, where $\alpha \geq 0$ is an exponent that represents the CSIT quality.
This power relationship can be used to express a variety of imperfect CSIT scenarios \cite{Caire2010}.
For example, $\alpha =0$ represents the case where the error power is fixed over the whole SNR regime, e.g. constant number of feedback bits.
On the other hand, $\alpha = \infty$ corresponds to perfect CSIT.
In DoF analysis, it is customary to truncate the exponent such that $\alpha \! \leq \! 1$, where $\alpha \! = \! 1$ is considered perfect from a DoF perspective as the error impact becomes negligible for $P_{t} \! \rightarrow \! \infty$.
Under this assumption, conventional MU transmission that employs Zero Forcing (ZF) Beamforming (BF) achieves a sum DoF of $K\alpha$.
Interestingly, the same sum DoF can be achieved using only part of the power given as $P_{t}^{\alpha}$.
This can be explained as follows:
in the presence of CSIT uncertainty, perfect interference nulling cannot be achieved, and any DoF gain achieved by increasing the power above a certain threshold (i.e. $P_{t}^{\alpha}$) is nullified by the increased interference.
Nevertheless, the remaining part (i.e. $P_{t}-P_{t}^{\alpha}$) can be used to transmit a common message which achieves a DoF of $1-\alpha$ by decoding it while treating the private messages as noise.
As the common message is decodable by all users, it can be cancelled from their received signals, and the DoF achieved by the private messages is unaffected.

\emph{Contribution and Organization}:
We employ JMB to maximize the minimum AR among users, where a given user's AR consists of a private AR, in addition to a portion of the common AR.
This is achieved by jointly optimizing the precoders and the partition factors which determine how much of the common message is allocated to each user.
This non-convex problem is transformed into an equivalent augmented Average Weighted Mean Square Error (AWMSE) problem, which is solved using Alternating Optimization (AO).
Simulation results show the superiority of the proposed scheme over conventional MU transmission, which does not incorporate the common message. For example, while conventional transmission with $\alpha = 0$ yields flat rates in the high SNR regime \cite{Joudeh2014}, JMB employs the multicast part to achieve a non-vanishing DoF.
To the best of our knowledge, multicast assisted robust fairness-based designs have not appeared in literature.

The rest of the paper is organized as follows:
the system model and problem formulation are given in Section \ref{Section_System_Model}.
The equivalent AWMSE problem is formulated in Section \ref{Section_ASR_AWSMSE_Optimization}. The AO algorithm is proposed in Section \ref{Section_AO}. Simulation results are given in Section \ref{Section_Numerical_Results}, and Section \ref{Section_conclusion} concludes the paper.

\emph{Notation}: Boldface uppercase letters denote matrices, boldface lowercase letters denote column vectors and standard letters denote scalars. $(\cdot)^{T}$ and $(\cdot)^{H}$ denote transpose and conjugate-transpose (Hermitian) operators, respectively. $\mathrm{tr}(\cdot)$ and $\|\cdot\|$ are the trace and Euclidian norm operators, respectively. $\mathrm{E}_{x}\{\cdot\}$ is the expectation w.r.t the random variable $x$.
%
\section{System Model and Problem Formulation}
\label{Section_System_Model}
For the MISO setup described in the previous section, the vector of complex data symbols is given as $\mathbf{s} \triangleq [s_{\mathrm{c}},s_{1},\ldots,s_{K}]^{T} \in\mathbb{C}^{K+1}$, where
$s_{\mathrm{c}}$ is the common symbol decodable by all users, $s_{i}$ is a private symbol intended solely for the $i$th user, $i\in\mathcal{K}$ and $\mathcal{K} \triangleq  \{1,\ldots,K\}$.
Entries of $\mathbf{s}$ have zero-means, unity powers and are mutually uncorrelated such that $\mathrm{E}\{\mathbf{s}\mathbf{s}^{H}\}=\mathbf{I}$.
$\mathbf{s}$ is linearly precoded as
\vspace{-2.0mm}
\begin{equation}\label{Eq_x}
  \mathbf{x} = \mathbf{p}_{\mathrm{c}}s_{\mathrm{c}} + \sum_{i=1}^{K}\mathbf{p}_{i}s_{i}
\end{equation}
%
where $\mathbf{x} \in\mathbb{C}^{N_{t}}$ is the transmit vector. $\mathbf{p}_{\mathrm{c}}\in\mathbb{C}^{N_{t}}$ and $\mathbf{p}_{i}\in\mathbb{C}^{N_{t}}$ are precoders for the common symbol and the $i$th private symbol respectively, from which
$\mathbf{P} \! \triangleq \!  \big[ \! \mathbf{p}_{\mathrm{c}},\mathbf{p}_{1},\ldots,\mathbf{p}_{K} \! \big]$ is composed.
The BS transmit power constraint is given as
$\mathrm{E}\{ \! \mathbf{x}^{H}\mathbf{x} \! \} \! = \! \mathrm{tr}\big( \! \mathbf{P}\mathbf{P}^{H} \! \big) \! \leq \! P_{t}$.
The $k$th received signal is given as
\vspace{-2.0mm}
\begin{equation}
\label{Eq_yk}
  y_{k} = \mathbf{h}_{k}^{H}\mathbf{x}+n_{k}
\end{equation}
where $\mathbf{h}_{k} \in \mathbb{C}^{N_{t}}$ is the channel impulse response vector between the BS and the $k$th user, from which the composite channel is defined as $\mathbf{H} \triangleq [\mathbf{h}_{1},\ldots,\mathbf{h}_{K}]$.
$n_{k} \thicksim \mathcal{CN} ( 0 , \sigma^{2}_{n_{k}} )$ is the AWGN. Throughout the paper, it is assumed that noise variances are equal across all users, i.e. $\sigma_{n_{k}}^{2}=\sigma_{n}^{2}, \ \forall k \in \mathcal{K}$.
\subsection{CSIT Uncertainty}
Each of the $K$ links exhibits independent fading, and remains almost constant over a frame of symbols, enabling users to estimate their channel vectors with high accuracy.
On the other hand, CSIT experiences uncertainty arising from limited feedback, delays or mismatches.
$\mathbf{H}$ is written as a sum of the transmitter-side channel estimate
$\widehat{\mathbf{H}}\triangleq[\widehat{\mathbf{h}}_{1},\ldots,\widehat{\mathbf{h}}_{K}]$
and the channel estimation error
$\widetilde{\mathbf{H}}\triangleq [\widetilde{\mathbf{h}}_{1},\ldots,\widetilde{\mathbf{h}}_{K}]$,
such that
%
$ \mathbf{H}= \widehat{\mathbf{H}} + \widetilde{\mathbf{H}}$.
%
The CSIT consists of $\widehat{\mathbf{H}}$, in addition to some statistical knowledge of $\widetilde{\mathbf{H}}$.
Particularly, the BS knows the probability distribution of the actual channel given the available estimate, i.e.
$f_{\mathbf{H}|\widehat{\mathbf{H}}}(\mathbf{H})=f_{\widetilde{\mathbf{H}}}(\mathbf{H}-\widehat{\mathbf{H}})$.
%
\subsection{MSE and MMSE}
\label{Subsection_MSE_MMSE_Rate}
The $k$th user obtains an estimate of the common symbol by applying a scalar equalizer $g_{\mathrm{c},k}(\mathbf{h}_{k})$ to
\eqref{Eq_yk} such that $\widehat{s}_{\mathrm{c},k}=g_{\mathrm{c},k}(\mathbf{h}_{k}) y_{k}$.
Assuming that the common symbol is successfully decoded by all users, the common symbol's receive signal part is reconstructed and cancelled from $y_{k}$. This improves the detectability of $s_{k}$, which is then estimated by applying $g_{k}(\mathbf{h}_{k})$ such that
$\widehat{s}_{k}=g_{k}(\mathbf{h}_{k}) (y_{k}-\mathbf{h}_{k}^{H}\mathbf{p}_{\mathrm{c}}s_{\mathrm{c},k})$.
The notations $g_{\mathrm{c},k}(\mathbf{h}_{k})$ and $g_{k}(\mathbf{h}_{k})$ are used to emphasise the dependencies on the actual channel, as each user is assumed to have perfect knowledge of its own channel vector.
$(\mathbf{h}_{k})$ is omitted for brevity unless special emphasis is necessary. This is also used with other variables that depend on the actual channel.
For the $k$th user, the MSEs  defined as $\varepsilon_{\mathrm{c},k} \triangleq \mathrm{E}_{\mathbf{s},n_{k}}\{|\widehat{s}_{\mathrm{c},k} - s_{\mathrm{c}}|^{2}\}$ and $\varepsilon_{k} \triangleq \mathrm{E}_{\mathbf{s},n_{k}}\{|\widehat{s}_{k} - s_{k}|^{2}\}$ are given as
\vspace{-2.0mm}
\begin{subequations}
\label{Eq_MSE}
\begin{align}
  \label{Eq_MSE_c_k}
  \varepsilon_{\mathrm{c},k}(\mathbf{h}_{k}) =& \ |g_{\mathrm{c},k}|^{2} T_{\mathrm{c},k} -2\Re \big\{g_{\mathrm{c},k}\mathbf{h}_{k}^{H}\mathbf{p}_{\mathrm{c}}\big\}+1 \\
  \label{Eq_MSE_k}
  \varepsilon_{k}(\mathbf{h}_{k}) =& \ |g_{k}|^{2} T_{k}-2\Re \big\{g_{k}\mathbf{h}_{k}^{H}\mathbf{p}_{k}\big\}+1
\end{align}
\end{subequations}
where $T_{\mathrm{c},k} \! = \! |\mathbf{p}_{\mathrm{c}}^{H}\mathbf{h}_{k}|^{2} \! + \!T_{k}$
and $T_{k} \! = \! \sum_{i=1}^{K} \! |\mathbf{p}_{i}^{H}\mathbf{h}_{k}|^{2} \! + \! \sigma_{n}^{2}$.
Optimum $g_{\mathrm{c},k}$ and $g_{k}$ are obtained from
$\frac{\partial \varepsilon_{\mathrm{c},k} }{\partial g_{\mathrm{c},k}} \! = \! 0 $ and
$\frac{\partial \varepsilon_{k} }{\partial g_{k}} \! = \! 0 $,
yielding the well-known Minimum MSE (MMSE) equalizers:
\vspace{-1.0mm}
\begin{equation}
 \label{Eq_g_MMSE}
  g_{\mathrm{c},k}^{\mathrm{MMSE}} \! (\mathbf{h}_{k}) \! = \! \mathbf{p}_{\mathrm{c}}^{H}\mathbf{h}_{k} T_{\mathrm{c},k}^{-1}
  \ \text{and} \
  g_{k}^{\mathrm{MMSE}} \! (\mathbf{h}_{k}) \! = \! \mathbf{p}_{k}^{H}\mathbf{h}_{k}T_{k}^{-1}.
\end{equation}
Substituting (\ref{Eq_g_MMSE}) into (\ref{Eq_MSE}), the $k$th user's MMSEs are given as
\vspace{-1.0mm}
\begin{equation}
  \label{Eq_MMSE}
  \varepsilon_{\mathrm{c},k}^{\mathrm{MMSE}}(\mathbf{h}_{k}) = T_{\mathrm{c},k}^{-1} E_{\mathrm{c},k}
  \quad \text{and} \quad
  \varepsilon_{k}^{\mathrm{MMSE}}(\mathbf{h}_{k}) = T_{k}^{-1}E_{k}
\end{equation}
where $E_{\mathrm{c},k} =  T_{\mathrm{c},k} - |\mathbf{p}_{\mathrm{c}}^{H}\mathbf{h}_{k}|^{2} = T_{k}$ and
$E_{k} =  T_{k} - |\mathbf{p}_{k}^{H}\mathbf{h}_{k}|^{2}$.
\subsection{Achievable Rate and Average Rate}
\label{subsection_A_Rates}
The MMSE and the Signal to Interference plus Noise Ratio (SINR) are related such that $\gamma_{\mathrm{c},k}={(1-\varepsilon_{\mathrm{c},k}^{\mathrm{MMSE}})}/{\varepsilon_{\mathrm{c},k}^{\mathrm{MMSE}}}$ and $\gamma_{k}={(1-\varepsilon_{k}^{\mathrm{MMSE}})}/{\varepsilon_{k}^{\mathrm{MMSE}}}$, where $\gamma_{\mathrm{c},k}$ and $\gamma_{k}$ are the $k$th user's SINRs.
Therefore, the $k$th user's maximum achievable common rate and private rate are written as $R_{\mathrm{c},k}\!(\mathbf{h}_{k}) \! = \! -\log_{2}(\varepsilon_{\mathrm{c},k}^{\mathrm{MMSE}})$ and $R_{k}\!(\mathbf{h}_{k}) \! = \! -\log_{2}(\varepsilon_{k}^{\mathrm{MMSE}})$, respectively.
The common message is transmitted at a common rate defined as $R_{\mathrm{c}} \triangleq \min_{j}\{R_{\mathrm{c},j}\}_{j=1}^{K}$, which ensures that it is decodable by all users.
As the multicast part is used to boost the individual rates achieved by users, the $k$th user is allocated a fraction $c_{k}$ of the common message, where $c_{k}\geq0$ and $\sum_{k=1}^{K}c_{k} = 1$.
The corresponding portion of the common rate is given as $c_{k}R_{\mathrm{c}}$.
In this case, the $k$th user's total achievable rate is given as $R_{k} + c_{k}R_{\mathrm{c}}$.
In a block-based transmission, this can be achieved by dedicating different parts of the common data block to different users, where any fraction $c_{k}$ can be achieved for sufficiently long blocks.
To achieve fairness among users, the objective would be to design $\mathbf{P}$ and $\mathbf{c}\triangleq[c_{1},\ldots,c_{K}]^{T}$ that maximize the minimum total rate among users defined as $R \triangleq \min_{k} \big\{ R_{k} + c_{k}R_{\mathrm{c}} \big\}_{k=1}^{K}$.
However, rates depend on the actual channel, and hence cannot be considered as optimization metrics at the BS.
Alternatively, we consider the Average Rates (ARs) defined as: $\mathrm{E}_{\mathbf{h}_{k}\mid \widehat{\mathbf{h}}_{k}}\{R_{\mathrm{c},k}\}$ and $\mathrm{E}_{\mathbf{h}_{k}\mid \widehat{\mathbf{h}}_{k}}\{R_{k}\}$, where averaging is taken over the CSIT error.
In the following, $\mathrm{E}_{\mathbf{h}_{k}\mid \widehat{\mathbf{h}}_{k}}\{\cdot\}$ will be simply referred to as $\mathrm{E}\{\cdot\}$.
%
\subsection{Sample Average Function}
In order to formulate a deterministic problem, the stochastic ARs are approximate by corresponding Sample Average Functions (SAFs)
obtained  by taking the ensemble average over a sample of $M$ independent identically distributed (i.i.d) realizations drawn from the distribution $f_{\mathbf{H}|\widehat{\mathbf{H}}}$.
The sample is defined as
$\mathbf{H}_{\mathcal{M}} \triangleq \left\{ \mathbf{H}^{(m)} \mid m \in \mathcal{M} \right\}$,
where $\mathbf{H}^{(m)} \triangleq [\mathbf{h}_{1}^{(m)} ,\ldots,\mathbf{h}_{K}^{(m)}]$ is the $m$th Monte-Carlo realization, and $\mathcal{M} \triangleq \left\{1,\ldots,M\right\}$.
The SAFs are given as: $\bar{R}_{\mathrm{c},k}^{(M)} = \frac{1}{M} \sum_{m=1}^{M} R_{\mathrm{c},k}^{(m)} $ and $\bar{R}_{k}^{(M)} =\frac{1}{M} \sum_{m=1}^{M} R_{k}^{(m)}$, where
$R_{\mathrm{c},k}^{(m)} \triangleq R_{\mathrm{c},k}\big(\mathbf{h}_{k}^{(m)}\big) $ and
$R_{k}^{(m)} \triangleq R_{k}\big(\mathbf{h}_{k}^{(m)}\big)$
are the rates associated with the realization $\mathbf{h}_{k}^{(m)}$.
In the following, the superscript $(m)$ is used to indicate the association of variables with the
$m$th Monte-Carlo realization.
It should be noted that $\mathbf{P}$ is fixed over the $M$ realizations of the rates, which follows from the definition of the ARs.
This also reflects the fact that $\mathbf{P}$ is optimized at the BS using partial CSI knowledge.
\newtheorem{Assumption_SNR}[Assumption_Counter]{Assumption}
\begin{Assumption_SNR}
\label{Assumption_SNR}
\textnormal{
In the following, we assume that $\frac{\sigma_{n}^{2}}{\|\mathbf{h}_{k}\|^{2}P_{t}} > 0$ with probability 1, $\forall k \in \mathcal{K}$.
}
\end{Assumption_SNR}
Alternatively, we can say that $\mathrm{SNR} = {P_{t}}/{\sigma_{n}^{2}}$ can only grow finitely large, and channel gains are finite.
Assumption \ref{Assumption_SNR} yields $\varepsilon_{\mathrm{c},k}^{\mathrm{MMSE}},\varepsilon_{k}^{\mathrm{MMSE}} > 0$ with probability $1$, as the presence of a nonzero noise variance dictates that $E_{\mathrm{c},k},E_{k}>0$.
This also implies that rates are finite, and by the strong law of large numbers we can write
\vspace{-2.0mm}
\begin{subequations}
\label{Eq_R_LLN}
\begin{align}
\label{Eq_R_LLN_c}
  \bar{R}_{\mathrm{c},k} \triangleq & \  \lim_{M\rightarrow \infty} \bar{R}_{\mathrm{c},k}^{(M)} = \mathrm{E} \{ R_{\mathrm{c},k} \},
  \text{almost surely} \\
  \label{Eq_R_LLN_p}
  \bar{R}_{k} \triangleq & \  \lim_{M\rightarrow \infty} \bar{R}_{k}^{(M)} = \mathrm{E} \{ R_{k} \},
  \text{almost surely}
\end{align}
\end{subequations}
where $\bar{R}_{\mathrm{c},k}$ and $\bar{R}_{k}$ are the approximated ARs for a sufficiently large $M$, which will be simply referred to as the ARs. Moreover, the common AR is defined as $\bar{R}_{\mathrm{c}} \triangleq \min_{j}\{\bar{R}_{\mathrm{c},j}\}_{j=1}^{K}$.
\subsection{Problem Formulation}
The minimum total AR among the $K$ users is defined as
$\bar{R} \triangleq \min_{k} \big\{ \bar{R}_{k} + c_{k}\bar{R}_{\mathrm{c}} \big\}_{k=1}^{K}$.
The objective is to design $\mathbf{P}$ and $\mathbf{c}$ such that  $\bar{R}$ is maximized.
This problem is formulated as
%
\vspace{-1.0mm}
\begin{subequations}
 \label{Eq_Opt_AR_M}
\begin{align}
 \label{Eq_Opt_AR_M_a}
\bm{\mathcal{R}}: &
\underset{\bar{R},\bar{R}_{\mathrm{c}}, \mathbf{P}, \mathbf{c} }{\max} \ \bar{R} \\
 \label{Eq_Opt_AR_M_b}
 \text{s.t.} \ &
  \bar{R}_{k} + c_{k} \bar{R}_{\mathrm{c}} \geq  \bar{R}, \; \forall k\in\mathcal{K} \\
 \label{Eq_Opt_AR_M_c}
  & \bar{R}_{\mathrm{c},k} \geq  \bar{R}_{\mathrm{c}}, \; \forall k\in\mathcal{K}  \\
 \label{Eq_Opt_AR_M_d}
   & \ c_{k} \geq 0, \; \forall k\in\mathcal{K}  \\
 \label{Eq_Opt_AR_M_e}
  & \sum_{k=1}^{K} c_{k} = 1 \\
 \label{Eq_Opt_AR_M_f}
  &   \mathrm{tr}\big(\mathbf{P}\mathbf{P}^{H}\big) \leq P_{t}
\end{align}
\end{subequations}
where the constraints in  \eqref{Eq_Opt_AR_M_b} and \eqref{Eq_Opt_AR_M_c} are introduced to eliminate the potential non-smoothness arising from the pointwise minimizations in $\bar{R}_{\mathrm{c}}$ and $\bar{R}$.
$\bm{\mathcal{R}}$ is non-convex and appears to be very challenging to solve.
\section{Equivalent AWMSE Optimization Problem}
\label{Section_ASR_AWSMSE_Optimization}
In this section, $\bm{\mathcal{R}}$ is transformed into a more tractable equivalent problem by exploiting the relationship between rates and augmented WMSEs, an approach inspired from \cite{Christensen2008,Razaviyayn2013b}.
We start by introducing the main components used to construct the equivalent problem, i.e. the augmented WMSEs:
\vspace{-2.0mm}
\begin{subequations}
\vspace{-2.0mm}
\label{Eq_A_WMSEs}
\begin{align}
\label{Eq_A_WMSEs_c}
\xi_{\mathrm{c},k}\big( \! \mathbf{h}_{k},g_{\mathrm{c},k},u_{\mathrm{c},k} \! \big) & =  u_{\mathrm{c},k}(\mathbf{h}_{k})\varepsilon_{\mathrm{c},k}(\mathbf{h}_{k}) \! - \! \log_{2} \! \big( \! u_{\mathrm{c},k}(\mathbf{h}_{k}) \! \big) \\
\label{Eq_A_WMSEs_p}
\xi_{k}\big(\mathbf{h}_{k},g_{k},u_{k}\big) & =  u_{k}(\mathbf{h}_{k})\varepsilon_{k}(\mathbf{h}_{k}) - \log_{2}\big(u_{k}(\mathbf{h}_{k})\big)
\end{align}
\end{subequations}
where $u_{\mathrm{c},k}(\mathbf{h}_{k})\geq 0$ and $ u_{k}(\mathbf{h}_{k}) \geq 0$ are weights associated with the $k$th user's MSEs. The dependencies of $\xi_{\mathrm{c},k}$ and $\xi_{k}$ on different variables are highlighted in \eqref{Eq_A_WMSEs} for their significance in the following analysis, where we establish the following WMSE-Rate relationship:
\vspace{-1.0mm}
\begin{equation}
\label{Eq_min_WMSE}
 \underset{u_{\mathrm{c},k}, g_{\mathrm{c},k}}{\min} \xi_{\mathrm{c},k} = 1-R_{\mathrm{c},k}
 \quad \text{and} \quad
 \underset{u_{k}, g_{k}}{\min} \ \xi_{k}= 1-R_{k}.
\end{equation}
This can be shown as follows. From
$\frac{\partial\xi_{\mathrm{c},k}}{\partial g_{\mathrm{c},k}} = 0$
and
$\frac{\partial\xi_{k}}{\partial g_{k}} = 0$, the optimum equalizers are obtained as
$g_{\mathrm{c},k}^{\ast}  = g_{\mathrm{c},k}^{\mathrm{MMSE}} $ and $g_{k}^{\ast} = g_{k}^{\mathrm{MMSE}}$.
Substituting this back into \eqref{Eq_A_WMSEs}, we obtain the augmented WMMSEs written as
\vspace{-1.0mm}
\begin{subequations}
\label{Eq_A_WMMSEs}
\begin{align}
\xi_{\mathrm{c},k}^{\mathrm{MMSE}}( \mathbf{h}_{k},u_{\mathrm{c},k}  \big) & =  u_{\mathrm{c},k}\varepsilon_{\mathrm{c},k}^{\mathrm{MMSE}} - \log_{2}(u_{\mathrm{c},k}) \\
\xi_{k}^{\mathrm{MMSE}}( \mathbf{h}_{k},u_{k} \big) & =  u_{k}\varepsilon_{k}^{\mathrm{MMSE}} - \log_{2}(u_{k}).
\end{align}
\end{subequations}
Furthermore, from
$\frac{\partial\xi_{\mathrm{c},k}^{\mathrm{MMSE}}}{\partial u_{\mathrm{c},k}} = 0$
and
$\frac{\partial\xi_{k}^{\mathrm{MMSE}}}{\partial u_{k}} = 0$, we obtain the optimum MMSE wights:
$u_{\mathrm{c},k}^{\ast} = u_{\mathrm{c},k}^{\mathrm{MMSE}} \triangleq \big( \varepsilon_{\mathrm{c},k}^{\mathrm{MMSE}} \big)^{-1}$
and
$u_{k}^{\ast} = u_{k}^{\mathrm{MMSE}} \triangleq \big( \varepsilon_{k}^{\mathrm{MMSE}} \big)^{-1}$.
Substituting this back into \eqref{Eq_A_WMMSEs} yields the relationship in \eqref{Eq_min_WMSE}.
It is evident from \eqref{Eq_MMSE} that the MMSE weights are dependent on the channel.

The equivalent problem is formulated using the augmented AWMSEs defined as:
$\mathrm{E}\{\xi_{\mathrm{c},k}\} $ and $\mathrm{E}\{\xi_{k}\}$.
Before we proceed, the augmented AWMSEs are approximated as:
%
\vspace{-2.0mm}
%
\begin{align}
\nonumber
\bar{\xi}_{\mathrm{c},k}^{(M)} & =  \frac{1}{M} \sum_{m=1}^{M}\xi_{\mathrm{c},k}^{(m)}
\quad \text{and} \quad
\bar{\xi}_{k}^{(M)} = \frac{1}{M} \sum_{m=1}^{M}\xi_{k}^{(m)}, \ \text{where} \\
\nonumber
\xi_{\mathrm{c},k}^{(m)} \! & \triangleq  \xi_{\mathrm{c},k}\big(  \mathbf{h}_{k}^{(m)} \! \!,g_{\mathrm{c},k}^{(m)} \! \!,u_{\mathrm{c},k}^{(m)}  \big)
\ \text{and} \
\xi_{k}^{(m)} \! \triangleq  \xi_{k}\big(  \mathbf{h}_{k}^{(m)} \! \!,g_{k}^{(m)} \! \!,u_{k}^{(m)}  \big)
\end{align}
correspond to the $m$th realization of the augmented WMSEs,
which depend on the $m$th realization of the equalizers:
$g_{\mathrm{c},k}^{(m)} \!  \triangleq \!  g_{\mathrm{c},k}\big( \!  \mathbf{h}_{k}^{(m)} \!  \big)  $
and
$g_{k}^{(m)} \!  \triangleq \!  g_{k}\big( \!  \mathbf{h}_{k}^{(m)} \!  \big)  $,
and the weights:
$u_{\mathrm{c},k}^{(m)} \!  \triangleq \!  u_{\mathrm{c},k}\big( \!  \mathbf{h}_{k}^{(m)} \!  \big)  $
and
$u_{k}^{(m)} \!  \triangleq \!  u_{k}\big(\! \mathbf{h}_{k}^{(m)} \!  \big)  $.
For compactness, we define the set of equalizers associated with the $M$ realizations and the $K$ users as:
$\mathbf{G} \!  \triangleq \!   \big\{ \!  \mathbf{g}_{\mathrm{c},k},\mathbf{g}_{k} \!  \mid \!  k \!  \in \!  \mathcal{K} \!  \big\}$,
where
$\mathbf{g}_{\mathrm{c},k} \!  \triangleq \!  \big\{ \!  g_{\mathrm{c},k}^{(m)} \!  \mid \!  m \!  \in \!  \mathcal{M} \!  \big\}$
and
$\mathbf{g}_{k} \!  \triangleq \!  \big\{ \!  g_{k}^{(m)} \!  \mid \!  m  \! \in \!  \mathcal{M} \!  \big\}$.
In a similar manner, we define:
$\mathbf{U} \!  \triangleq \!   \big\{ \!  \mathbf{u}_{\mathrm{c},k},\mathbf{u}_{k} \!  \mid \!  k \!  \in \!  \mathcal{K} \!  \big\}$,
where
$\mathbf{u}_{\mathrm{c},k} \!  \triangleq \!  \big\{ \!  u_{\mathrm{c},k}^{(m)} \!  \mid \!  m \!  \in \!  \mathcal{M} \!  \big\}$
and
$\mathbf{u}_{k} \!  \triangleq \!  \big\{ \!  u_{k}^{(m)} \!  \mid \!  m \!  \in \!  \mathcal{M} \!  \big\}$.
The approximated augmented AWMSEs for a sufficiently large $M$ are defined as
$\bar{\xi}_{\mathrm{c},k} \!  \triangleq \!  \lim_{M\rightarrow \infty} \!  \bar{\xi}_{\mathrm{c},k}^{(M)}$
and
$\bar{\xi}_{k} \!  \triangleq \!  \lim_{M\rightarrow \infty} \!  \bar{\xi}_{k}^{(M)}$,
which will be simply referred to as the AWMSEs.
The same approach used to prove \eqref{Eq_min_WMSE} can be employed to show that
\vspace{-1.0mm}
\begin{equation}
\label{Eq_min_AWMSE}
 \underset{\mathbf{u}_{\mathrm{c},k}, \mathbf{g}_{\mathrm{c},k}}{\min} \bar{\xi}_{\mathrm{c},k} = 1-\bar{R}_{\mathrm{c},k}
 \quad \text{and} \quad
 \underset{\mathbf{u}_{k}, \mathbf{g}_{k}}{\min} \ \bar{\xi}_{k}= 1-\bar{R}_{k}.
\end{equation}
where optimality conditions are checked separately for each realization.
The sets of optimum MMSE equalizers associated with \eqref{Eq_min_AWMSE} are defined as
$\mathbf{g}^{\mathrm{MMSE}}_{\mathrm{c},k} \! \triangleq \! \big\{ \! g_{\mathrm{c},k}^{\mathrm{MMSE}(m)}  \! \mid \! m \!  \in \!  \mathcal{M} \! \big\}$
and
$\mathbf{g}^{\mathrm{MMSE}}_{k} \! \triangleq \! \big\{ \! g_{k}^{\mathrm{MMSE}(m)} \! \mid \! m \!  \in \!  \mathcal{M} \!  \big\}$.
In the same manner, the sets of optimum MMSE weights are defined as
$\mathbf{u}^{\mathrm{MMSE}(m)}_{\mathrm{c},k} \! \triangleq \! \big\{ \! u_{\mathrm{c},k}^{\mathrm{MMSE}(m)} \! \mid \! m \!  \in \!  \mathcal{M} \! \big\}$
and
$\mathbf{u}^{\mathrm{MMSE}(m)}_{k}  \! \triangleq \! \big\{ \! u_{k}^{\mathrm{MMSE}(m)} \! \mid \! m \!  \in \!  \mathcal{M} \! \big\}$.
For the $K$ users, the MMSE solution is composed as
$\mathbf{G}^{\mathrm{MMSE}} \! \triangleq \!  \big\{ \! \mathbf{g}_{\mathrm{c},k}^{\mathrm{MMSE}}, \mathbf{g}_{k}^{\mathrm{MMSE}} \! \mid \! k \!  \in \!  \mathcal{K} \! \big\}$
and
$\mathbf{U}^{\mathrm{MMSE}} \! \triangleq  \! \big\{ \! \mathbf{u}_{\mathrm{c},k}^{\mathrm{MMSE}}, \mathbf{u}_{k}^{\mathrm{MMSE}} \! \mid \! k  \! \in \!  \mathcal{K} \! \big\}$.
\subsection{Augmented AWMSE Minimization}
Prompted by the relationship in \eqref{Eq_min_AWMSE}, the augmented AWMSE minimization problem is formulated as
\vspace{-2.0mm}
\begin{subequations}
 \label{Eq_Opt_AWMSE_M}
\begin{align}
 \label{Eq_Opt_AWMSE_M_a}
\bm{\mathcal{A}}: &
\underset{\bar{\xi},\bar{\xi}_{\mathrm{c}}, \mathbf{P}, \mathbf{c}, \mathbf{U}, \mathbf{G}}{\min} \ \bar{\xi} \\
 \label{Eq_Opt_AWMSE_M_b}
 \text{s.t.} \ &
  \bar{\xi}_{k} + c_{k} \big(\bar{\xi}_{\mathrm{c}} - 1 \big) \leq  \bar{\xi}, \; \forall k\in\mathcal{K} \\
 \label{Eq_Opt_AWMSE_M_c}
  & \bar{\xi}_{\mathrm{c},k} \leq  \bar{\xi}_{\mathrm{c}}, \; \forall k\in\mathcal{K}  \\
  \label{Eq_Opt_ASR_M_e}
  & \ c_{k} \geq 0, \; \forall k\in\mathcal{K}  \\
 \label{Eq_Opt_AWMSE_M_d}
  & \sum_{k=1}^{K} c_{k} = 1 \\
 \label{Eq_Opt_ASR_M_f}
  &   \mathrm{tr}\big(\mathbf{P}\mathbf{P}^{H}\big) \leq P_{t}
\end{align}
\end{subequations}
%
where $(\bar{\xi},\bar{\xi}_{\mathrm{c}})$ are auxiliary variables.
The equivalence between $\bm{\mathcal{A}}$ under the MMSE solution, and $\bm{\mathcal{R}}$ is demonstrated as follows.
Substituting
$\big(\mathbf{g}_{k}^{\mathrm{MMSE}},\mathbf{u}_{k}^{\mathrm{MMSE}}\big)$ and
$\big(\mathbf{g}_{\mathrm{c},k}^{\mathrm{MMSE}},\mathbf{u}_{\mathrm{c},k}^{\mathrm{MMSE}}\big)$
into \eqref{Eq_Opt_AWMSE_M_b}  and \eqref{Eq_Opt_AWMSE_M_c} respectively, and rearranging, we obtain
\vspace{-2.0mm}
\begin{subequations}
\label{Eq_R_AWMSE}
\begin{align}
\label{Eq_R_AWMSE_p}
\bar{R}_{k} +  c_{k} \big( 1 - \bar{\xi}_{\mathrm{c}} \big) \geq  &  1 - \bar{\xi}, \; \forall k\in\mathcal{K} \\
\label{Eq_R_AWMSE_c}
\bar{R}_{\mathrm{c},k} \geq &  1 - \bar{\xi}_{\mathrm{c}}, \; \forall k\in\mathcal{K}.
\end{align}
\end{subequations}
It can be seen that \eqref{Eq_R_AWMSE_p} and  \eqref{Eq_R_AWMSE_c} are equivalent to \eqref{Eq_Opt_AR_M_b} and \eqref{Eq_Opt_AR_M_c} respectively, where
\vspace{-1.0mm}
\begin{equation}
\label{Eq_R_AWMSE_slack}
\bar{R} = 1 - \bar{\xi} \quad \text{and} \quad \bar{R}_{\mathrm{c}}= 1 - \bar{\xi}_{\mathrm{c}}.
\end{equation}
Moreover, the monotonicity of the relationships in \eqref{Eq_R_AWMSE_slack} implies that the two problems are equivalent under the MMSE solution, i.e. maximizing $\bar{R}$ is equivalent to minimizing $\bar{\xi}$.
This relationship is exploited in the following section.
\section{Alternating Optimization Algorithm}
\label{Section_AO}
Although $\bm{\mathcal{A}}$ is non-convex in the joint set of optimization variables, it is convex  in each of the blocks
$\mathbf{G}$, $\mathbf{U}$, $\mathbf{c}$ and $\mathbf{P}$.
We propose an AO algorithm that exploits this block-wise convexity.
Each iteration of the algorithm consists of three steps: 1) updating $\mathbf{G}$ and $\mathbf{U}$, 2) updating $\mathbf{c}$, 3) updating $\mathbf{P}$.
\subsection{Updating the Equalizers and Weights}
\label{subsection_obj_fn_update}
In $n$th iteration of the AO algorithm, the equalizers and weights are updated as:
$\mathbf{G} = \mathbf{G}^{\mathrm{MMSE}}\big(\ddot{\mathbf{P}}\big)$ and $\mathbf{U} = \mathbf{U}^{\mathrm{MMSE}}\big(\ddot{\mathbf{P}}\big)$,
where $\ddot{\mathbf{P}}$ is the precoding matrix obtained from $(n-1)$th iteration.
For the resulting point $\big(\ddot{\mathbf{P}},\mathbf{G}^{\mathrm{MMSE}}(\ddot{\mathbf{P}}),\mathbf{U}^{\mathrm{MMSE}}(\ddot{\mathbf{P}})\big)$, the updated AWMSEs in \eqref{Eq_Opt_AWMSE_M} are denoted by $\ddot{\bar{\xi}}_{\mathrm{c},k}$ and $\ddot{\bar{\xi}}_{k}$. Moreover, the common AWMSE is given as $\ddot{\bar{\xi}}_{\mathrm{c}} = \min_{j} \{\ddot{\bar{\xi}}_{\mathrm{c},j}\}_{j=1}^{K}$. Finally, the cost function in \eqref{Eq_Opt_AWMSE_M} is reduced by minimizing the individual AWMSEs.
\subsection{Updating Partition Coefficients}
\label{Subsection_Opt_Split}
Using the updated AWMSEs from the previous subsection, and employing the relationships in \eqref{Eq_R_AWMSE} and \eqref{Eq_R_AWMSE_slack}, the problem of optimizing $\mathbf{c}$ is formulated in terms of the ARs as
%
\vspace{-2.0mm}
\begin{subequations}
 \label{Eq_Opt_C}
\begin{align}
 \label{Eq_Opt_C_a}
\bm{\mathcal{A}}_{\mathrm{c}}: &
\underset{\bar{R},\mathbf{c} }{\min} \ -\bar{R} \\
 \label{Eq_Opt_C_b}
 \text{s.t.} \ &
   \ddot{\bar{R}}_{k} + c_{k} \ddot{\bar{R}}_{\mathrm{c}}   \geq  \bar{R}, \; \forall k\in\mathcal{K} \\
 \label{Eq_Opt_C_c}
  & \sum_{k=1}^{K} c_{k} = 1 \\
 \label{Eq_Opt_C_d}
  & \ c_{k} \geq 0, \; \forall k\in\mathcal{K}
\end{align}
\end{subequations}
where $\ddot{\bar{R}}_{k} = 1-\ddot{\bar{\xi}}_{k} $ and $\ddot{\bar{R}}_{\mathrm{c}} = 1- \ddot{\bar{\xi}}_{\mathrm{c}} $.
While optimizing $\mathbf{c}$ does not influence the private ARs or the common AR, it redistributes the common AR among users in a way that further reduces the cost function of $\bm{\mathcal{A}}$.
A solution for problem $\bm{\mathcal{A}}_{\mathrm{c}}$ is developed by examining the KKT optimality conditions.
The Lagrangian of \eqref{Eq_Opt_C} is given as
\vspace{-2.0mm}
\begin{equation}
\nonumber
L =  -\bar{R} + \sum_{k=1}^{K} \mu_{k} \big( \bar{R} - \ddot{\bar{R}}_{k}  -  c_{k} \ddot{\bar{R}}_{\mathrm{c}}  \big)
+ \lambda \big( \sum_{k=1}^{K}c_{k} - 1 \big) - \sum_{k=1}^{K} \nu_{k} c_{k}
\end{equation}
where $\bm{\mu} \triangleq \{ \mu_{k} \mid k \in \mathcal{K}  \}$, $\lambda$, and $\bm{\nu} \triangleq \{ \nu_{k} \mid k \in \mathcal{K}  \}$ are the multipliers associated with the constraints in \eqref{Eq_Opt_C_a}, \eqref{Eq_Opt_C_b}, and \eqref{Eq_Opt_C_c}, respectively. The KKT conditions can be written as
\vspace{-2.0mm}
\begin{subequations}
\label{KKT_C}
\begin{align}
\label{KKT_C_a}
\frac{\partial L}{ \partial \bar{R}} &
= 0 \ \Rightarrow \sum_{k=1}^{K} \mu_{k} = 1 \\
\label{KKT_C_b}
\frac{\partial L}{ \partial c_{k}} &
= 0 \ \Rightarrow  -\mu_{k}\ddot{\bar{R}}_{\mathrm{c}} + \lambda - \nu_{k} = 0, \ \forall k \in \mathcal{K} \\
\label{KKT_C_c}
\mu_{k} & \geq 0, \
\mu_{k} \big( \bar{R} - \ddot{\bar{R}}_{k}  -  c_{k} \ddot{\bar{R}}_{\mathrm{c}}  \big) = 0, \ \forall k \in \mathcal{K} \\
\label{KKT_C_d}
\nu_{k} & \geq  0, \
\nu_{k} c_{k} = 0, \ \forall k \in \mathcal{K}.
\end{align}
\end{subequations}
For $\ddot{\bar{R}}_{\mathrm{c}}  = 0$, optimizing $\mathbf{c}$ is irrelevant.
On the other hand, $\ddot{\bar{R}}_{\mathrm{c}}  > 0$ implies that $\lambda>0$. This is concluded from \eqref{KKT_C_b} which can be rewritten as $\lambda = \nu_{k} + \mu_{k}\ddot{\bar{R}}_{\mathrm{c}}$, \eqref{KKT_C_a} which dictates that at least one element in $\bm{\mu}$ is nonzero, and the positivity of $\nu_{k}$ from \eqref{KKT_C_d}.
Now we consider $c_{k}$ under three cases:

1)
For the first case, we assume $\bar{R} < \ddot{\bar{R}}_{k}$. In this case,  \eqref{KKT_C_c} dictates that $\mu_{k} = 0$, as
$\bar{R} - \ddot{\bar{R}}_{k}  - c_{k} \ddot{\bar{R}}_{\mathrm{c}} = 0$ is impossible due to the positivity of $c_{k} \ddot{\bar{R}}_{\mathrm{c}} $.
Moreover, we have $\lambda = \nu_{k} > 0$ from \eqref{KKT_C_b}.
This requires $c_{k} = 0$ as seen from \eqref{KKT_C_d}.
In other words, if a given user's private AR is higher than the optimum $\bar{R}$, it will not be allocated a fraction of the common AR. Moreover, the corresponding constraints in \eqref{Eq_Opt_C_b} are inactive.

2)
Next, we consider $\bar{R} \! = \! \ddot{\bar{R}}_{k}$. Assuming that $c_{k} \! >  \!0$, then we must have $\mu_{k} \! = \! 0$ to satisfy \eqref{KKT_C_c}. This implies that $\lambda \! =  \! \nu_{k} \! > \! 0$, which contradicts \eqref{KKT_C_d}. Therefore, we must have $c_{k} \! = \! 0$ and $\bar{R} \! = \! \ddot{\bar{R}}_{k}$, i.e. the corresponding constraints in \eqref{Eq_Opt_C_b} are active.

3)
Finally, we consider $\bar{R} > \ddot{\bar{R}}_{k}$. Assuming that $\mu_{k} = 0$, then we have $\lambda = \nu_{k} > 0$ and $c_{k} = 0$. However, this is an infeasible solution as it contradicts \eqref{Eq_Opt_C_b}. Therefore, we must have $\mu_{k} > 0$ and
$\bar{R}  = \ddot{\bar{R}}_{k}  + c_{k} \ddot{\bar{R}}_{\mathrm{c}}$. For this case, the corresponding constraints in \eqref{Eq_Opt_C_b} are also active.

From the previous analysis, we can write
\vspace{-2.0mm}
\begin{equation}
\label{Eq_c_k}
c_{k} = \max \big\{0, \ddot{\bar{R}}_{\mathrm{c}}^{-1}(\bar{R} -  \ddot{\bar{R}}_{k}) \big\}, \ \forall k \in \mathcal{K}
\end{equation}
which suggests that the optimum $\mathbf{c}$ can be calculated using a form of water-filling, where $\bar{R}$ is the water level.
Without loss of generality, we assume that the private ARs are ordered in an ascending manner, i.e.
$\ddot{\bar{R}}_{i} \geq \ddot{\bar{R}}_{j} $, $\forall i,j\in {K}$ and $i > j$.
Assuming that all constraints in \eqref{Eq_Opt_C_b} are active, and using the constraint in \eqref{Eq_Opt_C_c}, the water level is calculated as
$\bar{R} = K^{-1} \big( \ddot{\bar{R}}_{\mathrm{c}} + \sum_{k=1}^{K} \ddot{\bar{R}}_{k} \big) $, from which we obtain
$c_{k} = \ddot{\bar{R}}_{\mathrm{c}}^{-1}(\bar{R} -  \ddot{\bar{R}}_{k}), \ \forall k \in \mathcal{K}$.
If the $K$th user (with the greatest private AR) has $c_{K} < 0$, then the first case of the three applies. The $K$th user is discarded and $\bar{R}$ and $\mathbf{c}$ are recalculated for the remaining $K-1$ users. Otherwise, the optimum solution is obtained.
This is summarized in Algorithm \ref{Algthm_C}.
\vspace{-2.0mm}
\begin{algorithm}
\caption{Optimizing the partition coefficients.}
\label{Algthm_C}
\begin{algorithmic}[1]
\State \textbf{Initialize}: $K_{\mathrm{c}} \gets K+1$
\Repeat
    \State $K_{\mathrm{c}} \gets K_{\mathrm{c}}-1$ and $\mathcal{K}_{\mathrm{c}} \gets \{1,\ldots,K_{\mathrm{c}}\}$
    \State $\bar{R} \gets K_{\mathrm{c}}^{-1}\big( \ddot{\bar{R}}_{\mathrm{c}} + \sum_{k=1}^{K_{\mathrm{c}}} \ddot{\bar{R}}_{k} \big) $
    \State $c_{k} \gets \ddot{\bar{R}}_{\mathrm{c}}^{-1} \big( \bar{R}  - \ddot{\bar{R}}_{k}  \big)$, $\forall k \in \mathcal{K}_{\mathrm{c}}$
    \State $c_{k} \gets 0$, $\forall k \in \mathcal{K} \setminus \mathcal{K}_{\mathrm{c}}$
\Until{$c_{K_{\mathrm{c}}} \geq 0$ }
\vspace{-1.0mm}
\end{algorithmic}
\end{algorithm}
\subsection{Updating the Precoders}
\label{Subsection_Opt Pre_Vect}
To formulate the problem of updating the precoders, the AWMSEs are written in terms of the updated blocks $\mathbf{G}$ and $\mathbf{U}$, and the block $\mathbf{P}$ which is yet to be updated.
For this purpose, we introduce the AWMMSE-components:
$\bar{\mathbf{\Psi}}_{\mathrm{c},k}$, $\bar{\mathbf{\Psi}}_{k}$, $\bar{t}_{\mathrm{c},k}$, $\bar{t}_{k}$,
$\bar{\mathbf{f}}_{\mathrm{c},k}$, $\bar{\mathbf{f}}_{k}$, $\bar{u}_{\mathrm{c},k}$, $\bar{u}_{k}$,
$\bar{\upsilon}_{\mathrm{c},k}$ and $\bar{\upsilon}_{k}$, which are obtained using the updated $\mathbf{G}$ and $\mathbf{U}$.
In particular, $\bar{u}_{\mathrm{c},k}$ and $\bar{u}_{k}$ are calculated by taking the ensemble averages over the $M$ realizations of
$u_{\mathrm{c},k}^{(m)}$ and $u_{k}^{(m)}$.
The rest are calculated in a similar manner by averaging over their corresponding realizations:
%
\vspace{-2.0mm}
\begin{align}
\nonumber
t_{\mathrm{c},k}^{(m)}& =  u_{\mathrm{c},k}^{(m)}\left|g_{\mathrm{c},k}^{(m)}\right|^{2}
&\text{and} &&
t_{k}^{(m)} & =  u_{k}^{(m)}\left|g_{k}^{(m)}\right|^{2}\\
\nonumber
\mathbf{\Psi}_{\mathrm{c},k}^{(m)} & =  t_{\mathrm{c},k}^{(m)} \mathbf{h}_{k}^{(m)}{\mathbf{h}_{k}^{(m)}}^{H}
&\text{and} &&
\mathbf{\Psi}_{k}^{(m)} & =   t_{k}^{(m)} \mathbf{h}_{k}^{(m)}{\mathbf{h}_{k}^{(m)}}^{H}\\
\nonumber
\mathbf{f}_{\mathrm{c},k}^{(m)} & = u_{\mathrm{c},k}^{(m)} \mathbf{h}_{k}^{(m)}{g_{\mathrm{c},k}^{(m)}}^{H}
&\text{and} &&
\mathbf{f}_{k}^{(m)} & =  u_{k}^{(m)} \mathbf{h}_{k}^{(m)}{g_{k}^{(m)}}^{H} \\
\nonumber
\upsilon_{\mathrm{c},k}^{(m)} & =  \log_{2}\left(u_{\mathrm{c},k}^{(m)}\right)
&\text{and} &&
\upsilon_{k}^{(m)} & =  \log_{2}\left(u_{k}^{(m)}\right).
\end{align}
Using the AWMMSE-components, the AWMSEs are given as
\vspace{-2.0mm}
\begin{subequations}
\vspace{-4.0mm}
\label{Eq_AWMSE}
\begin{align}
 \nonumber
  \bar{\xi}_{\mathrm{c},k}&  =  \mathbf{p}_{\mathrm{c}}^{H} \bar{\mathbf{\Psi}}_{\mathrm{c},k} \mathbf{p}_{\mathrm{c}}
    + \sum_{i=1}^{K}  \mathbf{p}_{i}^{H} \bar{\mathbf{\Psi}}_{\mathrm{c},k} \mathbf{p}_{i}
    + \sigma_{n}^{2}\bar{t}_{\mathrm{c},k}
    - 2\Re \big\{  \bar{\mathbf{f}}_{\mathrm{c},k}^{H} \mathbf{p}_{\mathrm{c}}  \big\}
    \\
    & \ +\bar{u}_{\mathrm{c},k} - \bar{\upsilon}_{\mathrm{c},k}
    \\
   \bar{\xi}_{k} &  =   \sum_{i=1}^{K}  \mathbf{p}_{i}^{H} \bar{\mathbf{\Psi}}_{k} \mathbf{p}_{i}
      +  \sigma_{n}^{2}\bar{t}_{k}
      -  2\Re \big\{ \bar{\mathbf{f}}_{k}^{H} \mathbf{p}_{k}\big\}
      +  \bar{u}_{k}   -   \bar{\upsilon}_{k}
\end{align}
\end{subequations}
from which the problem of optimizing $\mathbf{P}$  is formulated as
\vspace{-2.0mm}
\begin{subequations}
 \label{Eq_Opt_QCQP}
\begin{align}
\label{Eq_Opt_QCQP_a}
\bm{\mathcal{A}}_{\mathbf{P}}: &
\underset{\bar{\xi} , \bar{\xi}_{\mathrm{c}} ,  \mathbf{P}}{\min} \ \bar{\xi} \\
\nonumber
      \text{s.t.} \ &
    \sum_{i=1}^{K}  \mathbf{p}_{i}^{H} \bar{\mathbf{\Psi}}_{k} \mathbf{p}_{i}
    \! + \! \sigma_{n}^{2}\bar{t}_{k}
    \!- \! 2 \Re \big\{ \bar{\mathbf{f}}_{k}^{H} \mathbf{p}_{k} \big\}
    \! + \! \bar{u}_{k}
    \! -  \! \bar{\upsilon}_{k}
    \\
    \label{Eq_Opt_QCQP_b}
    &
    \! + \! c_{k}(\bar{\xi}_{\mathrm{c}} - 1)
    \leq \bar{\xi}, \ \forall k\in \mathcal{K} \\
\nonumber
&
     \mathbf{p}_{\mathrm{c}}^{H}
      \bar{\mathbf{\Psi}}_{\mathrm{c},k} \mathbf{p}_{\mathrm{c}}
    \! + \! \sum_{i=1}^{K}  \mathbf{p}_{i}^{H} \bar{\mathbf{\Psi}}_{\mathrm{c},k} \mathbf{p}_{i}
    \! + \! \sigma_{n}^{2}\bar{t}_{\mathrm{c},k}
    \!- \! 2 \Re \big\{ \bar{\mathbf{f}}_{\mathrm{c},k}^{H} \mathbf{p}_{\mathrm{c}} \big\}
    \\
     \label{Eq_Opt_QCQP_c}
     &
     +  \bar{u}_{\mathrm{c},k}
     -  \bar{\upsilon}_{\mathrm{c},k}
    \leq \bar{\xi}_{\mathrm{c}}, \ \forall k\in \mathcal{K} \\
 \label{Eq_Opt_QCQP_d}
                       &  \quad
                       \mathrm{tr}\big(\mathbf{P}\mathbf{P}^{H}\big) \leq P_{t}.
\end{align}
\end{subequations}
Problem \eqref{Eq_Opt_QCQP} is convex with quadratic constraints, which can be solved using off-the-self optimization software that employs interior-point methods \cite{Grant2008}.
\subsection{Alternating Optimization Algorithm}
\label{Subsection_AO}
The AO algorithm is constructed by repeating the steps described in the previous subsections until convergence.
This is summarized in Algorithm \ref{Algthm_AO} where $\epsilon_{R}$ determines the accuracy of the solution and $n_{\max}$ is the maximum number of iterations. Initializing $\mathbf{P}$ is discussed in Section \ref{Section_Numerical_Results}.
%
\vspace{-3.0mm}
\begin{algorithm}
\caption{Alternating Optimization}
\label{Algthm_AO}
\begin{algorithmic}[1]
\State \textbf{Initialize}: $n\gets 0$, $\hat{R}^{(n)} \gets 0$, $\mathbf{P}$
\label{Algthm_AO_step_initialize}
\Repeat
    \State \hspace{-17pt} $n\gets n+1$, $\ddot{\mathbf{P}}\gets \mathbf{P}$
    \State \hspace{-17pt} $\mathbf{G}\gets \mathbf{G}^{\mathrm{MMSE}}\big(\ddot{\mathbf{P}}\big)$, $\mathbf{U}\gets \mathbf{U}^{\mathrm{MMSE}}\big(\ddot{\mathbf{P}}\big)$
    \label{Algthm_AO_step_MMSE}
    \State \hspace{-17pt} $\mathbf{c} \gets \arg \bm{\mathcal{A}}_{\mathrm{c}}$
    \label{Algthm_AO_step_c}
    \State \hspace{-17pt} update
    $\big\{ \! \bar{\mathbf{\Psi}}_{\mathrm{c},k},\bar{\mathbf{\Psi}}_{k},\bar{\mathbf{F}}_{\mathrm{c},k},\bar{\mathbf{F}}_{k}, \bar{\mathbf{t}}_{\mathrm{c},k},\bar{\mathbf{t}}_{k},\bar{\mathbf{u}}_{\mathrm{c},k},\bar{\mathbf{u}}_{k},\bm{\bar{\upsilon}}_{\mathrm{c},k},\bm{\bar{\upsilon}}_{k}
    \! \big\}_{k\! = \! 1}^{K}$
    \State \hspace{-17pt}  $(\mathbf{P},\bar{\xi}^{(n)}) \gets \arg \bm{\mathcal{A}}_{\mathbf{P}}$ and $\hat{R}^{(n)} \gets 1- \bar{\xi}^{(n)}$
    \label{Algthm_AO_step_P}
\Until{$\left|\hat{R}^{(n)} - \hat{R}^{(n-1)} \right| < \epsilon_{R}$ \text{or} $n=n_{\max}$ }
\end{algorithmic}
\vspace{-1.0mm}
\end{algorithm}
\vspace{-4.0mm}
\newtheorem{Proposition_WAMSE_Conv}[Proposition_Counter]{Proposition}
\begin{Proposition_WAMSE_Conv}\label{Proposition_WAMSE_Conv}
\textnormal{
Algorithm \ref{Algthm_AO} monotonically increases the AR objective function of problem $\bm{\mathcal{R}}$ until convergence.
}
\end{Proposition_WAMSE_Conv}
\begin{proof}[Proof]
From the steps in Algorithm \ref{Algthm_AO}, we observe the sequence:
$1-\bar{\xi}^{(n-1)} \leq \ddot{\bar{R}} \leq \bar{R} \leq 1-\bar{\xi}^{(n)}$,
where $\ddot{\bar{R}} \! = \! \min_{k} \! \big\{\ddot{c}_{k}\ddot{\bar{R}}_{\mathrm{c}} \! + \! \ddot{\bar{R}}_{k}\big\}_{k=1}^{K}$ is the AR objective function at the output of step \ref{Algthm_AO_step_MMSE},
and
$\bar{R} \! = \! \min_{k} \! \big\{c_{k}\ddot{\bar{R}}_{\mathrm{c}} \! + \! \ddot{\bar{R}}_{k}\big\}_{k=1}^{K}$
is the AR objective function after updating $\mathbf{c}$ in step \ref{Algthm_AO_step_c}.
The relationships in \eqref{Eq_min_AWMSE}, \eqref{Eq_R_AWMSE} and \eqref{Eq_R_AWMSE_slack} hold, particularly at the output of steps \ref{Algthm_AO_step_MMSE} and \ref{Algthm_AO_step_c}.
This implies that each iteration increases the AR objective function. Moreover, the fact that $\bar{R}$ is bounded above for given $P_{t}$ ensures convergence. However, global optimality is not guaranteed due to the non-convexity of $\bm{\mathcal{R}}$.
\end{proof}
%
\section{Numerical Results}
\label{Section_Numerical_Results}
\begin{figure}[t!]\vspace{-4mm}
\centering
\includegraphics[width = 0.40\textwidth]{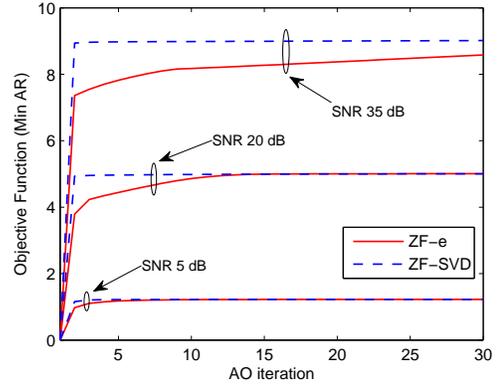}\\
\caption{Convergence of Algorithm \ref{Algthm_AO} using 1 randomly generated $\mathbf{H}$.}
\label{Fig_Convergence}
\vspace{-5mm}
\end{figure}
%
\begin{figure}[t!]\vspace{-3mm}
\centering
\includegraphics[width = 0.40\textwidth]{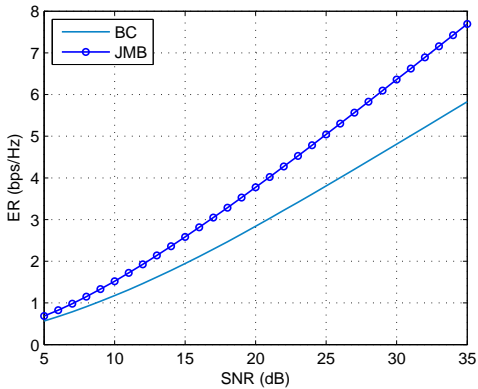}\\
\caption{MISO-BC and MISO-JMB ERs. $K=2$, and $\sigma_{e}^{2} = P_{t}^{-0.6}$. }
\label{Fig_ER_Beta_06}
\vspace{-3mm}
\end{figure}
%
%
\begin{figure}[t!]\vspace{-1mm}
\centering
\includegraphics[width = 0.40\textwidth]{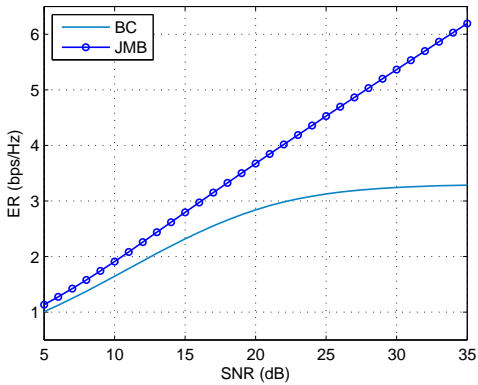}\\
\caption{MISO-BC and MISO-JMB ERs. $K=2$, and $\sigma_{e}^{2} = 0.063$. }
\label{Fig_ER_Beta_0}
\vspace{-5mm}
\end{figure}
%
We consider a system with $N_{t} \! = \! K \! = \! 2$ and uncorrelated channel fading, where the entries of $\mathbf{H}$ have a complex Gaussian distribution $\mathcal{C}\mathcal{N}\left(0,1\right)$.
The noise variance is fixed as $\sigma_{n}^{2} \! = \! 1$, yielding a long-term SNR of $P_{t}$.
Moreover, Gaussian CSIT error is assumed where the entries of $\widetilde{\mathbf{H}}$ are generated according to the distribution $\mathcal{C}\mathcal{N}\left(0,\sigma_{e}^{2}\right)$.
For each realization $\mathbf{H}$, a channel estimation error $\widetilde{\mathbf{H}}$ is drawn from $\mathcal{C}\mathcal{N}\left(0,\sigma_{e}^{2}\right)$, from which the channel estimate is calculated as
$\widehat{\mathbf{H}} \! = \! \mathbf{H} \! - \! \widetilde{\mathbf{H}}$.
A channel realization $\mathbf{H}$ should not be confused with a Monte-Carlo realization $\mathbf{H}^{(m)}$.
The former is the actual channel which is unknown to the BS, and assumed to remain constant for a given transmission.
On the other hand, the latter is part of a sample $\mathbf{H}_{\mathcal{M}}$ employed in the optimization at the BS.
The size of the sample is set to $M \!= \! 1000$.
For a given $\widehat{\mathbf{H}}$, the $m$th Monte-Carlo realization is obtained as
$\mathbf{H}^{(m)} \!  = \! \widehat{\mathbf{H}} \! + \!  \widetilde{\mathbf{H}}^{(m)}$, where $\widetilde{\mathbf{H}}^{(m)}$ is drawn from $\mathcal{C}\mathcal{N}\left(0,\sigma_{e}^{2}\right)$.

First, we examine the convergence of Algorithm \ref{Algthm_AO} using two different $\mathbf{P}$ initializations.
The first initialization (ZF-e) is taken as the sum-DoF motivated design in \cite{Hao2013}.
The private precoders are initialized as
$\mathbf{p}_{k} \! = \! \sqrt{P_{t}^{\alpha}/K} \widehat{\mathbf{p}}_{k}^{\mathrm{ZF}}$,
where $\widehat{\mathbf{p}}_{k}^{\mathrm{ZF}}$ is a normalized ZF-BF vector constructed using the channel estimate $\widehat{\mathbf{H}}$.
The common precoder is given as $\mathbf{p}_{\mathrm{c}} \! = \! \sqrt{P_{t} - P_{t}^{\alpha}}\mathbf{e}_{1}$, where $\mathbf{e}_{1}$ is a vector with $1$ as the first entry and zeros elsewhere.
The second initialization (ZF-SVD) retains the ZF-BF part and the power allocation. However, the common precoder is obtained as the dominant left singular vector of $\widehat{\mathbf{H}}$.
Figure \ref{Fig_Convergence} shows the AR convergence for $\sigma_{e}^{2} \! = \! P_{t}^{-0.6}$, and SNRs $5$, $20$ and $35$ dB.
It is evident that the algorithm eventually converges to a limit point regardless of the initialization.
Moreover, better convergence performance is achieved with ZF-SVD compared to ZF-e due to the initialization of the common precoder.
The following results employ ZF-SVD.

Next, we consider the Ergodic Rate (ER) performance.
For a given channel estimate $\widehat{\mathbf{H}}$ and the corresponding error statistics, maximizing the minimum AR yields the solution $\mathbf{P}$.
Employing $\mathbf{P}$ for the channel realization $\mathbf{H}$ yields the achievable rates defined in \ref{subsection_A_Rates},
from which the achievable minimum rate among users (i.e. $R$) is obtained.
The ER is defined as $\mathrm{E}_{\mathbf{H}} \! \left\{  R  \right\}$,
which captures the overall achievable performance for all possible channel realizations.
In the following simulations, the ER is calculated by averaging over $200$ randomly generated channel realizations.
Conventional fairness-based BC transmission is considered as a baseline, which is obtained by discarding the common ARs and $\mathbf{c}$ in $\bm{\mathcal{R}}$.
A decaying estimation error power of $\sigma_{e}^{2} \! = \! P_{t}^{-0.6}$ is used in Figure \ref{Fig_ER_Beta_06}.
On the other hand, Figure \ref{Fig_ER_Beta_0} uses a fixed estimation error power $\sigma_{e}^{2} = \!  \big(10^{\frac{20}{10}}\big)^{-0.6} \! \! \! = 0.063$, which is equivalent to the CSIT quality obtained at SNR 20 dB in the previous case.
It is evident that users benefit from the incorporation of the common message over the entire SNR range. The significance of this utility grows with increased SNR.
For example, Figure \ref{Fig_ER_Beta_06} is a manifestation of DoF gains translating into rate gains at finitely high SNRs.
Moreover, Figure \ref{Fig_ER_Beta_0} demonstrates JMB's virtue of achieving fairness with non-vanishing growth in a scenario where conventional transmission hits a performance ceiling.
%
\section{Conclusion}
\label{Section_conclusion}
A multicast assisted scheme was proposed to achieve max-min fairness in a MU-MISO system with partial CSIT and perfect CSIR.
Precoders are designed and the common message rate is divided among users, such that the minimum AR among users is maximized. This problem is transformed into an AWMSE problem solved using a converging AO algorithm.
Simulation results show that the proposed scheme achieves significant rate gains over conventional transmission, which does not incorporate the multicast part.
%
%
%
\ifCLASSOPTIONcaptionsoff
  \newpage
\fi
\bibliographystyle{IEEEtran}
\bibliography{IEEEabrv,References}
\end{document}